\documentclass{article}

\usepackage{axodraw,graphicx,color}

\usepackage[english]{babel}

\begin{document}

\title{{\bf  Relativistic heavy ion collider as a photon factory:
 from GDR excitations to vector meson production }}
\author{
I.A. Pshenichnov\footnote{e-mail: Pshenichnov@inr.troitsk.ru}
 \\  
{\em Institute for Nuclear Research, Russian Academy of Science,}\\
{\em 117312 Moscow, Russia}\\
}
\date{ }
\maketitle

\begin{abstract}

A variety of phenomena, which reveal itself in distant collisions
of ultrarelativistic nuclei is discussed. One or both nuclei may be 
disintegrated in a single collision event by the long-range electromagnetic
forces due to the impact of the Lorentz-boosted Coulomb fields of
collision partners.The process is considered in the framework of 
the Weizs\"{a}cker-Williams method and simulated by 
the RELDIS code, which takes into account all possible channels
of nuclear disintegration, including multiple neutron and proton emission and
meson production. Mutual electromagnetic dissociation of nuclei in 
peripheral collisions can best be studied at RHIC and LHC.
The contributions of next-to-leading-order processes with
multiple photon absorption of equivalent photons by collision
partners are considered in detail. As demonstrated, the 
rates of the correlated forward-backward $2n$ and $3n$ 
emission are very sensitive to the presence of 
double and triple excitations of Giant Resonances in colliding nuclei.
A practical application consists in the possibility of beam luminosity 
monitoring in colliders via the registration of the 
correlated $1n$ and $2n$ emission in mutual electromagnetic dissociation.

\end{abstract}

\section{Introduction}

The colliders of ultrarelativistic nuclei, the Relativistic Heavy Ion 
Collider (RHIC) at Brookhaven National Laboratory
and the future Large Hadron Collider (LHC) at CERN, 
were designed for the primary aim
to provide conditions for creating and detecting quark-gluon plasma.
The conditions to create such an unexplored state of matter are
expected to be fulfilled in central hadronic collisions of heavy nuclei 
like Au and Pb.

However, new phenomena in heavy ion collisions are anticipated 
also beyond the domain of impact parameter $b\leq R_1+R_2$, where hadronic 
interactions of nuclei with the nuclear radii $R_1$ and $R_2$ are confined.
Due to the long-range electromagnetic interaction, ultraperipheral collisions of
nuclei with $b\ge R_1+R_2$ can be also considered. According to 
the Weizs\"{a}cker-Williams method~\cite{Krauss:vr,Baur:2001jj}, 
the Lorentz-contracted Coulomb fields of moving charges can be represented as 
intense sources of equivalent virtual photons with a wide energy spectrum.
Depending on virtual photon energy, the absorption of photons leads to
a variety of photonuclear reactions with emission of nucleons, nuclear fragments
and mesons~\cite{Pshenichnov:1999hw}. This is commonly termed as electromagnetic
dissociation (ED) of nuclei. Low-energy few-MeV photons are the most 
frequent in the spectrum, and the excitation of Giant Resonances (GR) 
in nuclei, in particular, Giant Dipole Resonance (GDR) 
followed by neutron emission is the most probable
channel of electromagnetic dissociation.

\subsection{Mutual electromagnetic 
dissociation of heavy ions in colliders}

Since the first pioneering studies of the electromagnetic 
dissociation~\cite{Heckman,Olson}, the process has gained a wide-spread 
interpretation as disintegration of one of the collision
partners in ultraperipheral collisions of heavy ions without direct
overlap of their nuclear densities. 
This consideration is fully justified 
in fixed target experiments, where the electromagnetic dissociation 
of either projectile or target nuclei is detected. Very recent
experiments~\cite{Scheidenberger:tr,Hill:ie} can be mentioned as 
examples of projectile and target dissociation measurements,
respectively.   

Both at the RHIC and LHC colliders, the single electromagnetic dissociation 
cross section by far exceeds the geometrical hadronic cross section 
due to the direct nuclear overlap, as shown by 
calculations~\cite{Krauss:vr,Baur:2001jj,Baltz:as,Pshenichnov:2001qd}. 
As a result, electromagnetic dissociation  and 
$e^+e^-$-pair production followed by electron capture will reduce the beam 
life-time in the colliders~\cite{Baltz:as}.

The consideration of single electromagnetic dissociation should be 
extended further, since both of the colliding nuclei may be disintegrated in a single 
event by their Coulomb fields. In this case
mutual electromagnetic dissociation takes 
place~\cite{Pshenichnov:2001qd}. This process can best be studied in 
heavy-ion colliders, in contrast to fixed target experiments, and
has a practical application. It makes possible to monitor collider
luminosity via the registration of correlated forward-backward neutrons produced in 
mutual electromagnetic dissociation.

In addition, the experimental studies of the mutual electromagnetic
dissociation of heavy nuclei in colliders can provide valuable information 
on double and triple excitations of Giant Resonances in heavy nuclei. 
Such exotic collective excitations
of nuclear matter are difficult to study in fixed target experiments, since
the contributions of double and triple excitations are at the level of $\sim1$\% and
$\sim 0.01$\%, respectively~\cite{Llope:vp}.

\section{Calculation of electromagnetic dissociation cross sections}

According to the Weizs\"{a}cker-Williams 
method~\cite{Krauss:vr,Baur:2001jj}, 
the impact of the Lorentz-boosted Coulomb field of the nucleus $A_1$ 
is treated as the absorption of   
equivalent photons by the nucleus $A_2$.     

In the rest frame of this nucleus the spectrum of photons
from the collision partner $A_1$ at impact parameter $b$ is expressed as:
\begin{equation}
N_{Z_1}(E_1 ,b)=\frac{\alpha Z^{2}_1}{\pi ^2}
\frac{{\sf x}^2}{\beta ^2 E_1 b^2}
\Bigl(K^{2}_{1}({\sf x})+\frac{1}{\gamma ^2}K^{2}_{0}({\sf x})\Bigl).
\label{WWspectrum}
\end{equation}
Here $\alpha$ is the fine structure constant,
${\sf x}=E_1 b/(\gamma \beta \hbar c)$ is an argument of the modified Bessel
functions of zero and first orders, $K_0$ and $K_1$, $\beta =v/c$ and
$\gamma =(1-\beta^2)^{-1/2}$ is the Lorentz factor of the moving charge
$Z_1$ in the rest frame of $A_2$. If the Lorentz factor of each heavy-ion
beam is $\gamma_{beam}$ in the laboratory system, then
$\gamma= 2\gamma^2_{beam} -1$ for the case of collider.
For example, at LHC the Coulomb fields of ions are extremely 
Lorentz-contracted, $\gamma\sim 1.7\times 10^7$.
\begin{figure}[htb]
\begin{centering}
{\includegraphics[width=0.8\textwidth]{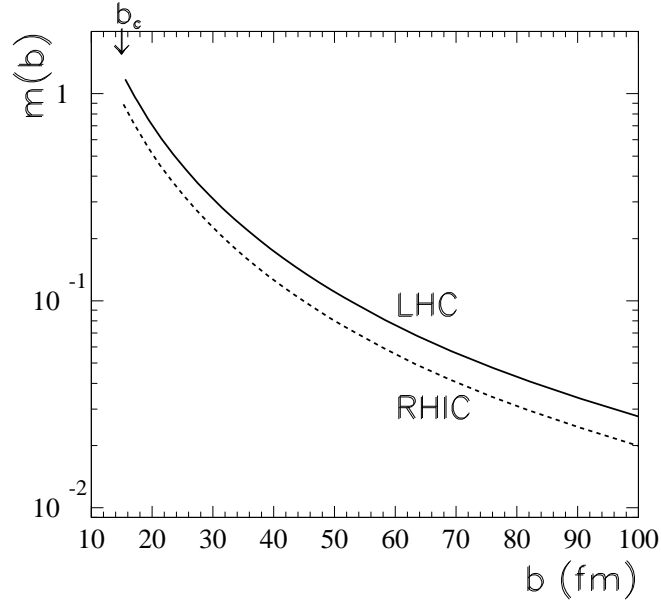}}
\caption{Mean number of photons absorbed in AuAu collisions at RHIC (dashed line) 
and in PbPb collisions at LHC (solid line) as a function of impact parameter.}
\label{m}
\end{centering}
\end{figure}

The mean number of photons absorbed by the nucleus $A_2$
in the collision at impact parameter $b$ is defined by:
\begin{equation}
m_{A_2}(b)=
\int\limits_{E_{min}}^{E_{max}}N_{Z_1}(E_1 ,b)\sigma_{A_2}(E_1)dE_1 ,
\label{mb}
\end{equation}
where the appropriate total photoabsorption cross section
$\sigma_{A_2}(E_1)$ is used. For $E_{min}$ one usually takes
the neutron emission threshold, while
the upper limit of integration is $E_{max}\approx\gamma/R_{1,2}$.
The calculation results for $m(b)$ are shown in Fig.~\ref{m} for
$100+100$~ A~GeV AuAu collisions at RHIC and for $2.75+2.75$~A~TeV PbPb 
collisions at LHC. The range of integration in Eq.~(\ref{mb})  
extends well above the GR region, which gives the main contribution.
However, the inverse-square dependence $m(b)\sim 1/b^2$ 
mentioned in Ref.~\cite{Baur:2001jj} still can be used to 
approximate the results shown in Fig.~\ref{m}.
The lower limit of impact parameter 
in electromagnetic interactions is approximately given by the
sum of nuclear radii, $b_{c}\approx R_1+R_2$.
Despite the fact that nuclei are relatively transparent
for photons, the mean number of photons absorbed in 
close electromagnetic collisions ($b\sim b_c$) 
is not small, $m(b)\sim 1$,
at RHIC and LHC. This is due to the fact that the flux of equivalent 
photons is quite high due to the extreme Lorentz 
contraction of Coulomb fields and high $Z$ of colliding nuclei.

\subsection{Single dissociation}

Assuming the Poisson distribution for the number of absorbed photons, the 
total single dissociation cross section for the nucleus 
$A_2$ due to the exchange of one
photon is given by:
\begin{equation}
\sigma^{ED}_{s}({\rm LO})=2\pi\int\limits_{b_{c}}^{\infty} bdb P_s(b)
=2\pi\int\limits_{b_{c}}^{\infty} bdb m_{A_2}(b)e^{-m_{A_2}(b)},
\label{S1}
\end{equation}
for the leading order (LO) process. Here the probability of single LO
dissociation $P_s(b)$ at given impact parameter $b$ is introduced.
Consequently, for the 
next-to-leading order (NLO) process with two exchanged photons (NLO$_2$):
\begin{equation}
\sigma^{ED}_{s}({\rm NLO_{2}})=2\pi\int\limits_{b_{c}}^{\infty} bdb 
\frac{m^2_{A_2}(b)}{2}e^{-m_{A_2}(b)}.
\label{S2}
\end{equation}

\subsection{Leading order process of mutual dissociation} 

\begin{figure}[tb]
\unitlength=1.0cm 
\begin{center}
\begin{picture}(7.0,6.0)(3.,-1.)
%%%%%%%%%%%%%%%%%%%%%%%%%%%%%%%% LO
\Text(2.5,3.8)[]{\Large {\bf LO}}  
\SetScale{0.8} 
\SetWidth{1.0}
\ArrowLine(0,100)(60,100)  
\Text(0,3.12)[]{$A_1$} 
\SetWidth{0.5}  
\GCirc(60,100){4}{1.} 
\SetWidth{1.0}
\ArrowLine(64,100)(120,100)  
\Text(2.48,3.12)[]{$A_1$}   
\SetWidth{0.5}
\Photon(60,97)(60,36){3}{4} 
\Text(1.36,1.76)[]{$E_1$} 
\SetWidth{1.0}
\ArrowLine(0,33)(60,33)  
\Text(0,0.64)[]{$A_2$} 
\SetWidth{0.5}  
\GCirc(60,33){4}{0} 
\SetWidth{2.0}
\ArrowLine(60,33)(120,33)  
\Text(2.48,0.64)[]{$A_2^\star$}
\SetWidth{0.5}  
\GCirc(120,100){4}{0}
\SetWidth{2.0} 
\ArrowLine(120,33)(180,33)  
\Text(4.72,0.64)[]{$A_2^\star$}
\SetWidth{0.5}  
\GCirc(120,33){4}{1.} 
\Photon(120,97)(120,36){3}{4} 
\Text(3.05,1.76)[]{$E_2$}
\SetWidth{2.0}
\ArrowLine(120,100)(180,100)  
\Text(4.72,3.12)[]{$A_1^\star$}
%%%%%%%%%%%%%%%%%%%%%%%%% NLO-12
\SetOffset(6.8,2.2)
\Text(2.5,3.5)[]{\large {\bf NLO$_{12}$}}    
\SetWidth{1.0}
\ArrowLine(0,100)(60,100)  
\Text(0.,3.05)[]{$A_1$} 
\SetWidth{0.5}  
\GCirc(60,100){4}{1.} 
\SetWidth{1.0}
\ArrowLine(64,100)(120,100)  
\Text(2.50,3.05)[]{$A_1$}   
\SetWidth{0.5}
\Photon(60,97)(60,36){3}{4} 
\Text(1.36,1.76)[]{$E_1$} 
\SetWidth{1.0}
\ArrowLine(0,33)(60,33)  
\Text(0,0.64)[]{$A_2$} 
\SetWidth{0.5}  
\GCirc(60,33){4}{0} 
\SetWidth{2.0}
\ArrowLine(60,33)(120,33)  
\Text(2.48,0.64)[]{$A_2^\star$}
\SetWidth{0.5}  
\GCirc(120,100){4}{0}
\SetWidth{2.0} 
\ArrowLine(120,33)(220,33)  
\Text(5.52,0.64)[]{$A_2^\star$}
\SetWidth{0.5}  
\GCirc(120,33){4}{1.} 
\Photon(120,97)(120,36){3}{4} 
\Text(3.05,1.76)[]{$E_2$}
\GCirc(140,100){4}{0}
\Photon(140,97)(140,36){3}{4} 
\Text(4.24,1.76)[]{$E_3$}
\GCirc(140,33){4}{1.}
\SetWidth{2.0}
\ArrowLine(120,100)(220,100)  
\Text(5.52,3.05)[]{$A_1^\star$}
%%%%%%%%%%%%%%%%%%%%%%%%%%% NLO-22
\SetOffset(6.8,-1.8)
\Text(2.5,3.5)[]{\large {\bf NLO$_{22}$}}    
\SetWidth{1.0}
\ArrowLine(-20,100)(60,100)  
\Text(0,3.05)[]{$A_1$} 
\SetWidth{0.5}  
\GCirc(60,100){4}{1.} 
\SetWidth{1.0}
\ArrowLine(64,100)(120,100)  
\Text(2.50,3.05)[]{$A_1$}   
\SetWidth{0.5}
\Photon(60,97)(60,36){3}{4}
\Text(0.88,1.76)[]{$E_1$} 
\Photon(40,97)(40,36){3}{4} 
\GCirc(40,100){4}{1.}
\GCirc(40,33){4}{0}
\Text(2.0,1.76)[]{$E_2$} 
\SetWidth{1.0}
\ArrowLine(-20,33)(60,33)  
\Text(0,0.64)[]{$A_2$} 
\SetWidth{0.5}  
\GCirc(60,33){4}{0} 
\SetWidth{2.0}
\ArrowLine(60,33)(120,33)
\Line(40,33)(60,33)  
\Text(2.48,0.64)[]{$A_2^\star$}
\SetWidth{0.5}  
\GCirc(120,100){4}{0}
\SetWidth{2.0} 
\ArrowLine(120,33)(220,33)  
\Text(5.52,0.64)[]{$A_2^\star$}
\SetWidth{0.5}  
\GCirc(120,33){4}{1.} 
\Photon(120,97)(120,36){3}{4} 
\Text(3.12,1.76)[]{$E_3$}
\GCirc(140,100){4}{0}
\Photon(140,97)(140,36){3}{4} 
\Text(4.24,1.76)[]{$E_4$}
\GCirc(140,33){4}{1.}
\SetWidth{2.0}
\ArrowLine(120,100)(220,100)  
\Text(5.52,3.05)[]{$A_1^\star$}   
\end{picture} 
\end{center} 
\caption{Mutual electromagnetic excitation of 
relativistic nuclei: leading order (LO) contribution, 
next-to-leading-order contribution with single and double photon
exchange processes (${\rm NLO_{12}}$) and next-to-leading-order 
contribution with two double photon exchange processes (${\rm NLO_{22}}$). 
Open and closed circles denote elastic and 
inelastic vertices, respectively.  
\label{fig:EMsecond}  
} 
\end{figure}
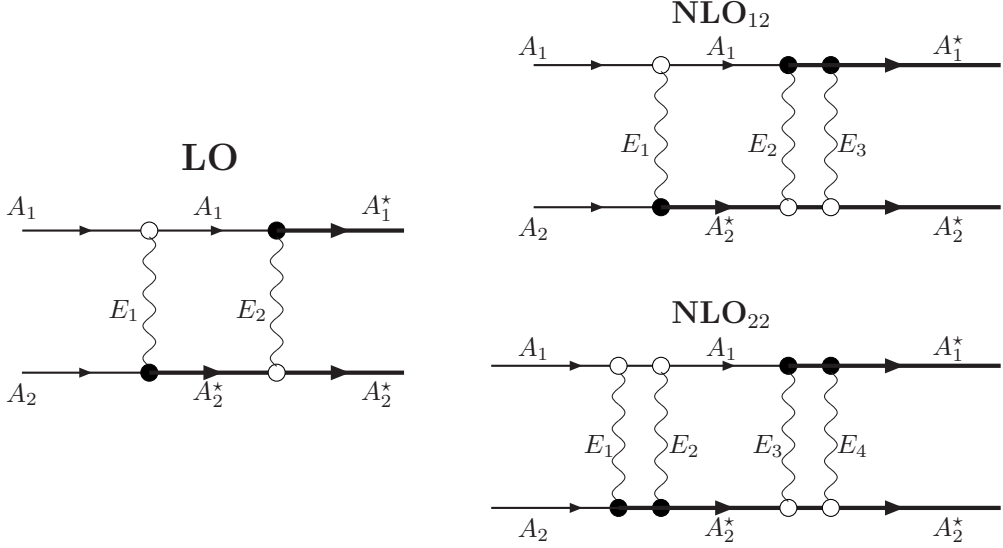

Following the assumption that the primary and secondary photon exchanges in
the LO process of mutual dissociation shown in Fig.~\ref{fig:EMsecond} 
may be considered independently (see details in Ref.~\cite{Pshenichnov:2001qd}),
one can write the cross section
for the dissociation of nuclei $A_1$ and $A_2$ to channels $i$ 
and $j$, respectively, as:
\begin{equation}
\sigma^{ED}_m(i\mid j)=2\pi\int\limits_{b_{c}}^{\infty} bdb 
P_{A_1}(b) P_{A_2}(b),
\label{eq:SM}
\end{equation}
In Eq.~(\ref{eq:SM})  the probability of dissociation of the nucleus $A_2$ 
at impact parameter $b$ via the channel $i$ is defined as:  
\begin{equation}
P_{A_2}(b)=e^{-m_{A_2}(b)}\int\limits_{E_{min}}^{E_{max}} dE_1 N_{Z_1}(E_1,b)
\sigma_{A_2}(E_1)f_{A_2}(E_1,i).
\label{eq:P1}
\end{equation}
$N_{Z_1}(E_1 ,b)$ is the spectrum of virtual photons 
from the collision partner $A_1$ at impact parameter $b$.
$\sigma_{A_2}(E_1)$ and $f_{A_2}(E_1,i)$ are the total photoabsorption cross 
section and the branching ratio for the channel $i$, respectively,
for the absorption of a photon with energy $E_1$ on the nucleus $A_2$. 
Naturally, the expression for $P_{A_1}(b)$ is obtained by exchange of
subscripts.

The total cross section for the mutual electromagnetic dissociation 
due to the leading order process shown in Fig.~\ref{fig:EMsecond} 
is given by:
\begin{equation}
\sigma^{ED}_m({\rm LO})=2\pi\int\limits_{b_{c}}^{\infty} bdb P_m(b)
=2\pi\int\limits_{b_{c}}^{\infty} bdb m_A^2(b) e^{-2m_A(b)},
 \label{eq:LO}
\end{equation}
\noindent 
where the mutual dissociation probability $P_m(b)$ at given impact parameter $b$
is introduced, and the case of equal masses and charges of 
collision partners is considered, $A_1=A_2=A$ and $Z_1=Z_2=Z$.

\subsection{Next-to-leading-order processes}

In addition to the leading order process of mutual dissociation, 
a set of next-to-leading order (NLO) processes with exchange of  
three or four photons should be considered.  
The total cross section for the process with three photons shown as  
${\rm NLO_{12}}$ in Fig.~\ref{fig:EMsecond} is given by:
\begin{equation}  
\sigma^{ED}_m({\rm NLO_{12}})=2\pi\int\limits_{b_{c}}^{\infty} bdb 
\frac{m_A^3(b)}{2} e^{-2m_A(b)}. 
\label{eq:NLO12}
\end{equation}
\begin{center}
\begin{table}[tb]
\caption{Total mutual electromagnetic  
dissociation cross sections for the leading order,
next-to-leading order processes and for the sum of all contributions
for Pb-Pb collisions at LHC.  
}
\vspace{0.2cm}
\begin{tabular}{|c|c|}
\hline
 & \\
Cross section &  2.75+2.75 A TeV Pb-Pb \\
(barns) & at LHC  \\
 &  \\
\hline\hline 
 &  \\
$\sigma^{ED}_m({\rm LO})$ & 3.92 \\
 &  \\
$\sigma^{ED}_m({\rm NLO_{12}})+\sigma^{ED}_m({\rm NLO_{21}})$ & 1.50 \\
 &  \\
$\sigma^{ED}_m({\rm NLO_{22}})$  & 0.23 \\
 &  \\
Triple excitations: NLO$_{\rm TR}$ & 0.56 \\
 &  \\
$\sigma^{ED}_m({\rm tot})$  &  6.21 \\
 &  \\
\hline
\end{tabular}
\label{tab:TLO-NLO}
\end{table}
\end{center}
\noindent A complementary process (${\rm NLO_{21}}$) with the excitation 
of the nucleus $A_2$ via double photon absorption is equally possible 
and has the same cross section.

Another next-to-leading order process of mutual dissociation is due 
to exchange of four photons (${\rm NLO_{22}}$ in 
Fig.~\ref{fig:EMsecond}), and its cross section can be written as:
\begin{equation}  
\sigma^{ED}_m({\rm NLO_{22}})=2\pi\int\limits_{b_{c}}^{\infty} bdb 
\frac{m_A^4(b)}{4} e^{-2m_A(b)}. 
\label{eq:NLO22}
\end{equation}
Calculations of  $\sigma^{ED}_m(i\mid j)$, $\sigma^{ED}_m({\rm LO})$, 
$\sigma^{ED}_m({\rm NLO})$ were 
performed by the modified code RELDIS~\cite{Pshenichnov:2001qd}, 
which contains a special 
simulation mode for the mutual electromagnetic dissociation process.
As one can see from Tab.~\ref{tab:TLO-NLO}, the LO mechanism
gives $\sim 63$\% of the $\sigma^{ED}_m({\rm tot})$ 
at LHC energies. The sum of the NLO contributions
to the total cross section gives additional
$\sim 28$\%. Therefore, as expected at LHC, the remaining contribution 
of 0.56 b  ($\sim 9$\% of the total mutual ED cross section)   
is due to exotic triple nuclear excitations (NLO$_{\rm TR}$) 
with three and more photons absorbed by {\em at least one } 
of the collision partners. This includes the following processes in our
notation: NLO$_{23}$, NLO$_{32}$, NLO$_{33}$, and also 
NLO$_{34}$, NLO$_{43}$, NLO$_{44}$

\section{Mutual dissociation as a filter to select close collisions and high-order
excitations}

\begin{figure}[htb]
\begin{centering}
{\includegraphics[width=1.05\textwidth]{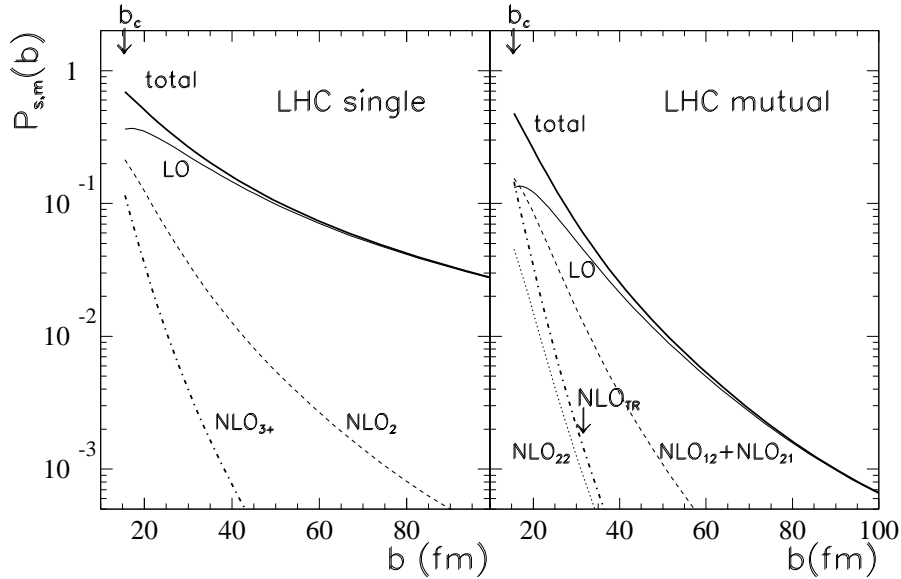}}
\caption{Probability of single (left panel) and mutual (right panel) 
LO and NLO ED processes as a function of impact parameter 
in PbPb collisions at LHC. Thick solid lines
present the sum of all LO and NLO contributions. }
\label{prob}
\end{centering}
\end{figure}

Most mutual ED events take place in close collisions with
small $b\geq b_c$, where the probability to absorb a virtual 
photon is large, and hence, two or more photons can be absorbed
by each of the collision partners.
This is demonstrated in Fig.~\ref{prob}, where the probabilities for single 
$P_s(b)$ and mutual $P_m(b)$ electromagnetic dissociation are shown as
functions of impact parameter $b$ for each of the LO and NLO processes.
The mutual dissociation probabilities $P_m(b)$ have much steep decrease 
as $b$ increases compared to $P_s(b)$.  In other words, 
detection of particles emitted in mutual dissociation
can be used as a filter to select close collisions.
The NLO contributions in mutual dissociation are noticeably enhanced compared to
single dissociation, as shown in Fig.~\ref{prob}. 
The sum of NLO$_{12}$ and NLO$_{21}$ contributions approaches the LO contribution
in the region of close collisions $b\sim b_c$.
In this region the probability of triple excitations NLO$_{\rm TR}$ is also 
found to be comparable with the LO contribution. 
However, all the NLO contributions have steeper 
decrease compared to the LO contribution, and 
the resulting NLO cross sections given in Tab.~\ref{tab:TLO-NLO} are found to be lower 
compared to the LO cross sections. 

As recently confirmed by calculations~\cite{Baltz:2002pp}, mutual 
electromagnetic excitation of nuclei can be also used to tag 
production of vector mesons,
$\rho^0, \omega, \phi$ and $J/\psi$ by virtual photons in collisions 
with smaller average $b$.

Double and triple excitations are expected to be frequent in mutual 
electromagnetic dissociation,
$\sim 28$\% and $\sim 9$\%, respectively, see Tab.~\ref{tab:TLO-NLO}.     
Therefore, the experimental studies of the mutual dissociation of heavy ions 
at LHC can provide valuable information on double and triple nuclear excitations
in electromagnetic interactions. 
This is particularly important for
the triple Giant Resonance excitations, since currently there 
are no experimental
data on such extreme excitations and only first theoretical predictions for
the positions and widths of such states were given recently~\cite{dePassos:2001dc}.

\section{Neutron emission in mutual dissociation}

The numbers of forward neutrons emitted in the mutual ED process and registered
by the Zero Degree Calorimeters~\cite{ZDC-ALICE02}  
can be used to obtain information on multiple 
excitations at LHC~\cite{Pshenichnov:ALICE}.
For example, if the decay channel of one of the collision partners 
in mutual dissociation is not exactly known, one can define 
inclusive mutual dissociation cross sections, $\sigma_m(1nX\mid {\cal D})$, 
$\sigma_m(2nX\mid {\cal D})$, $\sigma_m(3nX\mid {\cal D})$ for
emission of one, two and three neutrons, respectively, by the other partner.
In such notations ${\cal D}$ denotes an arbitrary 
dissociation mode, while $X$ denote 
any particle, except neutron. In the case, when the numbers of neutrons
are exactly known, one can define semi-inclusive mutual 
neutron emission cross sections $\sigma_m(1nX\mid 1nY)$, 
$\sigma_m(1nX\mid 2nY)$ and $\sigma_m(2nX\mid 2nY)$.

The cross sections for some specific channels of mutual dissociation 
at LHC predicted by the RELDIS model~\cite{Pshenichnov:2001qd} 
are given in Tab.~\ref{tab:TLONLO}. 
As one can note, $\sigma_m^{ED}(1nX\mid 1nY)$ has a small
NLO correction, $\sim 7$\%, while $\sigma_m^{ED}(3nX\mid D)$ becomes almost
twice as large as its LO value if NLO correction is included. 
This is due to the fact that
the NLO processes shown in Fig.~\ref{fig:EMsecond} include
nuclear excitation due to double photon absorption and, particularly,
the double GDR excitation process. Since the average GDR energy for Pb
is about 13--14 MeV, the double GDR excitation introduces, on average, 
26--28 MeV excitation energy which is already above the $3n$ emission threshold.
\begin{center}
\begin{table}[tb]
\caption{Mutual electromagnetic dissociation cross sections for
Pb-Pb collisions at LHC. $X$ and $Y$ denote any particle, except neutron.
${\cal D}$ means any dissociation channel. Calculation results are given (a) 
for the leading order contribution only, (b) for the sum of 
leading order and next-to-leading-order contributions.}
\vspace{0.3cm}
\begin{tabular}{|c|c|c|c|}
\hline
 & & & \\
 &Cross & (a) & (b) \\ 
 &section & LO & LO+NLO$_{12}$+ \\
 &(mb) & & NLO$_{21}$+NLO$_{22}$ \\
 & & & \\
\hline\hline
 & & & \\
 & $\sigma_m^{ED}(1nX\mid 1nY)$& 750 & 805 \\
 & & & \\
2.75+2.75 A TeV & $\sigma_m^{ED}(1nX\mid {\cal D})$& 1698 & 2107 \\
 & & & \\
Pb-Pb at LHC & $\sigma_m^{ED}(2nX\mid {\cal D})$& 443 & 654 \\
 & & & \\
 & $\sigma_m^{ED}(3nX\mid {\cal D})$& 241 & 465 \\
 & & & \\
\hline
\end{tabular}
\label{tab:TLONLO}
\end{table}
\end{center}
Therefore, as a rule, $1n$ and $2n$ emission cross sections are less
changed by taking into account NLO corrections than $3n$ cross sections. 
The measurements of the rates of forward $3n$ emission 
can be proposed to detect multiple GDR excitations in nuclei.

\subsection*{Acknowledgments}
It is a pleasure to thank Jakob Bondorf, Igor Mishustin and Alberto Ventura for
collaboration. The author thanks the organizers of the Seminar 
for a fruitful scientific atmosphere and a possibility to contribute
with this talk. 
This work was supported by the Russian Foundation for Basic Research, 
grant 02-02-16013.

%%%%%%%%%%%%%%%%%%%%%%%%%%%%%%%%
\end{document}